\RequirePackage{fix-cm}
\documentclass[smallextended]{svjour3}       
\smartqed  
\usepackage{graphicx}
\begin{document}

\title{Low-Energy Antinucleon-Nucleus Interaction Revisited}



\author{E. Friedman } 


\institute{E. Friedman \at
              Racah Institute of Physics, the Hebrew University,
              Jerusalem 91904, Israel \\
              Tel.: +972-2-6584667\\
              Fax: +972-2-6586347\\
              \email{elifried@cc.huji.ac.il}           
}

\date{Received: date / Accepted: date}

\maketitle

\begin{abstract}
Annihilation cross sections of antiprotons and antineutrons
on the proton between 50 and 400 MeV/c show Coulomb focusing below
200 MeV/c and almost no charge-dependence above 200 MeV/c. Similar
comparisons for heavier targets are not possible for lack of overlap
between nuclear targets studied with $\bar p$ and $\bar n$ beams.
Interpolating between $\bar p$-nucleus annihilation cross sections
with the help of an optical potential
to compare with  $\bar n$-nucleus annihilation cross sections reveal
unexpected features of Coulomb interactions in the latter.
Direct comparisons between $\bar n$-nucleus and $\bar p$-nucleus  
annihilations at very
low energies could be possible if $\bar p$ cross sections are measured
on the same targets and at the same energies as the available cross sections
for $\bar n$.  Such measurements may be feasible in the foreseeable future.

\end{abstract}

\section{Introduction}
\label{intro}
The motivation for the present work is the paper by Astrua et al.
\cite{ABB02} reporting annihilation cross sections for antineutrons
on six targets at momenta below 400 MeV/c. Very large cross sections 
were measured below 180 MeV/c but quantitative analyses  
have not been reported except noting that features of strong absorption
are observed in these cross sections.
Unfortunately direct comparisons with corresponsing cross sections
for antiproton-nucleus annihilation could not be made because 
experimental results are available mostly for
different targets at different energies.
In contrast, annihilation cross sections for antiprotons and 
antineutrons on the proton are available between 50 and 400 MeV/c,
showing the expected Coulomb effects at the lower end of this range
and hardly any dependence on charge at the higher end. 

\begin{table}
\caption{Experimental results for antinucleon-nucleus interaction
at low energies}
\label{tab:data}       
\begin{tabular}{lllll}
\hline\noalign{\smallskip}
target& $\bar p$ atoms&$\bar p$ ann.&$\bar p$ scatt.&$\bar n$ ann.\\ \hline
\noalign{\smallskip}\hline\noalign{\smallskip}
 C& +& &+&+\\
 O& +& & & \\
 Ne& & +& & \\
 Al& &  & & +\\
 Ca&+&  &+&  \\
 Fe&+& & &   \\
 Ni&+& +& &   \\
 Cu& &  & & + \\
 Zr&+&  & &   \\
 Ag& &  & & + \\
 Cd&+&  & &   \\
 Sn&+& +& & + \\
 Te&+&  & &   \\
 Pt& & +& &   \\
 Pb&+& &+&+ \\  \hline
 data points   & 90 & 7 & 88 & 42 \\
\noalign{\smallskip}\hline
\end{tabular}
\end{table}

Table \ref{tab:data} shows a schematic summary of available 
experimental results for low energy antinucleon interactions
with nuclei. 
It is seen that only for Sn there are results for both $\bar p$
and $\bar n$ annihilation cross sections, (but not at the same
energies).
In order to compare between $\bar n$ and $\bar p$ 
annihilation cross sections 
on nuclear targets, we use optical-model interpolations for $\bar p$.
The present work includes some updates compared to Ref. \cite{Fri14}
where more details are included. 

In Sect.~\ref{sec:ptarget} we compare total annihilation cross sections
for $\bar p p$  and $\bar n p$ at very low energies and describe
the mechanism of Coulomb focusing which is responsible for the 
large differences observed. In Sect.~\ref{sec:pbarsNUC}
we establish an optical-model approach to the $\bar p$-nucleus
interaction which is used in Sect.~\ref{sec:nbarsNUC} for comparisons
with the results of Ref. \cite{ABB02} for $\bar n$-nucleus cross
sections. In the summary (Sect.~\ref{sec:summ}) it is proposed to match 
the existing data
of annihilation cross sections
for $\bar n$ on nuclei by measuring annihilation cross sections for
$\bar p$ on the same targets at the corresponding energies, in an attempt
to shed light on what appears to be a puzzle.

\section{Antiproton and antineutron annihilation on the proton}
\label{sec:ptarget}

\begin{figure}
  \includegraphics[width=0.75\textwidth]{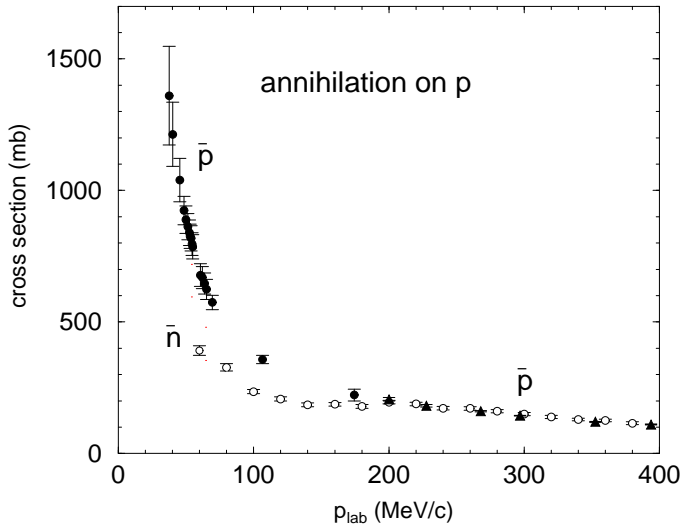}
\caption{Experimental annihilation cross sections
for $\bar p$ and $\bar n$ on the proton, see text for references.
Open circles for $\bar n$, filled circles and triangles for $\bar p$.}
\label{fig:pbarnbarp}      
\end{figure}

Figure \ref{fig:pbarnbarp} shows experimental annihilation cross sections
of $\bar p$ and $\bar n$ on the proton. All the $\bar n$ cross sections
and the $\bar p$ cross sections below 180 MeV/c, open circles and solid
circles, respectively, were measured by the
OBELIX collaboration \cite{BBC96,ZBB99,Ben97,BBC97} and the $\bar p$
cross sections above 180 MeV/c, solid triangles,
were measured by Br\"uckner et al.
\cite{BCD90} using a different experimental setup. 
The figure shows that the results in the two momentum ranges
join very smoothly and that above 200 MeV/c there are no observed
differences between the $\bar pp$ and $\bar np$ cross sections, although
the former contains contributions of I=0 and I=1 isospin states
whereas the latter is a pure I=1 isospin state.

The $\bar n$ cross sections increase
as the energy goes down, most likely due to the
expected $1/v$ dependence of the $s$-wave cross section. However, the increase
of the cross sections for the $\bar p$ is much stronger
than the increase for $\bar n$,
resulting from the Coulomb focusing effect which has 
been observed in annihilation cross sections of $\bar p$ on nuclei
\cite{BFG01}. For very strong absorption which is typical of 
antiproton interactions, the `black disk' cross section $\pi R^2$,
with $R$ the radius, is replaced, classically, by $\pi d_0^2$ 
where $d_0$ is the impact parameter for which the
distance of closest approach is R. A straight-forward calculation yields
for the total reaction cross section
\begin{equation}
\label{eq:CF}
\sigma_R=\pi R^2 (1+\frac{m+M}{M}\frac{Ze^2}{RE_{lab}})
\end{equation}
with $m$ and $M$ the masses of the $\bar p$ and target nucleus,
respectively, and $E_{lab}$ the lab. kinetic energy.
A quantum-mechanical calculation \cite{BFG01} leads to an identical
result for the Coulomb enhancement factor
\begin{equation}
\label{eq:QF}
\sigma_R=\pi R^2 (1+\frac{2mZe^2}{\hbar^2k_{lab} kR})
\end{equation}
with $k$ and $k_{lab}$ the cm and lab wave numbers,
respectively. It was shown in \cite{BFG01} that at very low energies
$\sigma _R \approx ZR \approx ZA^{1/3}$,
and for $R=1.84+1.12A^{1/3} $~fm good agreement with experiment
is obtained. 

\begin{table}
\caption{Energies and momenta where the Coulomb enhancement factor is 2}
\label{tab:Coul}
\begin{tabular}{llll}
\hline\noalign{\smallskip}
target  &R(fm) &$E_{lab}$(MeV) &$p_{lab}$(MeV/c) \\
\noalign{\smallskip}\hline\noalign{\smallskip}
p & 3.0 & 1.0 & 43 \\
C & 4.4 & 2.1 & 63 \\
Cu& 6.3 & 6.6 & 112 \\
Sn& 7.4 & 9.8 & 136 \\
Pb& 8.5 & 14  & 162 \\ 
\noalign{\smallskip}\hline
\end{tabular}
\end{table}

Table \ref{tab:Coul} shows calculated energies and momenta where
the Coulomb enhancement factor is 2. For a proton target that momentum is
near 45~MeV/c, as is indeed observed
in Fig.~\ref{fig:pbarnbarp}. It is therefore concluded that at very low energies
the major differences between $\bar p$ and $\bar n$ annihilation on the
proton are due to the Coulomb focusing effect.

\section{Antiproton-nucleus optical potential}
\label{sec:pbarsNUC}

Having demonstrated that the differences between 
low energy $\bar p$  and $\bar n$
annihilation cross sections on the proton are fully explained by the
Coulomb focusing effect, we turn to the $\bar n$ annihilation
cross sections on six nuclear targets between C and Pb \cite{ABB02}. 
Here it is
impossible to compare directly between experimental results and we 
use an optical potential to calculate cross sections for $\bar p$
on the same targets and at the same energies as for the 
$\bar n$ measurements. Only an outline is presented here as
more details are given in \cite{Fri14}.

For very low energies of interest we begin by analyzing the 
extended and precise
experimental results for strong interaction effects in antiprotonic
atoms, adopting the `global' analysis of 90 data points across the
periodic table done a decade ago \cite{FGM05}. The simplest
`$t \rho$' form  of the optical potential is  

\begin{equation}
\label{eq:pbarspotl}
2\mu V_{{\rm opt}}(r) = -4\pi(1+\frac{\mu}{m}
\frac{A-1}{A})[b_0(\rho_n+\rho_p)
  +b_1(\rho_n-\rho_p)]~~,
\end{equation}
where $\mu$ is the reduced mass of the $\bar p$,
 $\rho_n$ and $\rho_p$ are the neutron and proton density
distributions normalized to the number of neutrons $N$ and number
of protons $Z$, respectively, $A=N+Z$, and $m$ is the mass of the
nucleon. The complex parameters $b_0$ and $b_1$ are determined by
fits to the data. The isovector coefficient $b_1$ was shown in
\cite{FGM05} to be consistent with zero and as we found in the 
previous section that no isovector dependence is observed in the
annihilation cross sections on the proton, we set $b_1$=0 for the
rest of this work. Finite range folding of a $\bar p$-nucleon
interaction is also included \cite{FGM05}.

\begin{figure}
  \includegraphics[width=0.75\textwidth]{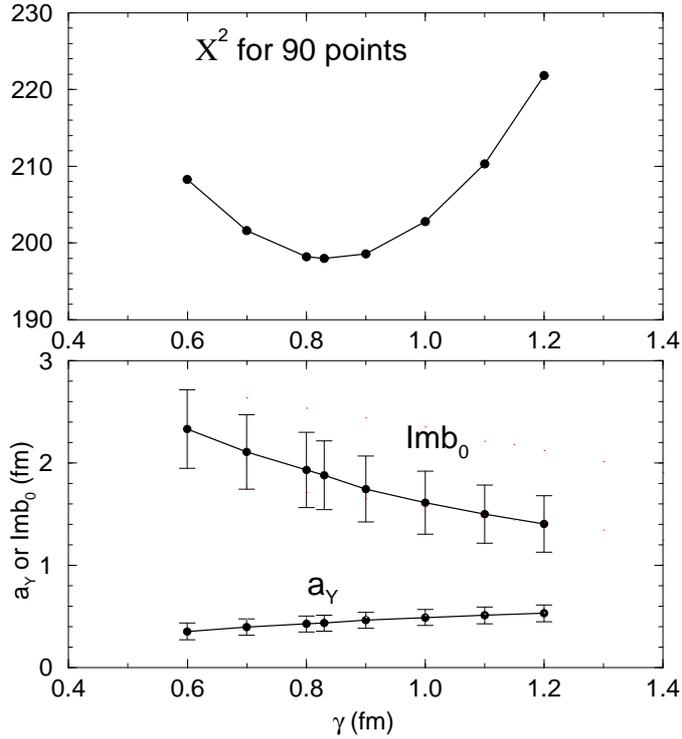}
\caption{Summary of global fits to $\bar p$ atoms as function of the
neutron density parameter $\gamma$ of Eq.(\ref{eq:RMF})
using a Yukawa-type $\bar p$-nucleon interaction. Top: $\chi ^2$
values for 90 data points, bottom: resulting parameters Im $b_0$ and $a_Y$,
see text.}
\label{fig:pbarsA}      
\end{figure}

Proton densities were obtained from experimentally determined charge
densities of nuclei. Neutron densities were approximated by
two-parameter Fermi distributions, characterized by the difference
between rms radii of neutron density and the corresponding 
proton density distribution in nuclei, parameterized by \cite{FGa07}
\begin{equation} \label{eq:RMF}
r_n-r_p = \gamma \frac{N-Z}{A} + \delta .
\end{equation}
Figure \ref{fig:pbarsA} shows examples of fits to the data for
a Yukawa-type $\bar p$-N interaction. The minimum in $\chi ^2$ is
obtained for a value of the parameter $\gamma $ which implies
values of $r_n-r_p$ in agreement with the great majority of results
regarding these differences, obtained by a variety of 
methods \cite{JTL04,Fri09,TWG14}. In
\cite{Fri14} similar results are presented for a Gaussian interaction.
It is worth noting that the rms radii involved and the strong 
interaction parameters $b$ are the same, within errors, for the two
interaction models.

The above optical potential was shown in \cite{Fri14} to reproduce 
very well angular distributions for elastic scattering of $\bar p$
by C, Ca and Pb at 300 MeV/c. We therefore proceed to calculate 
from this potential annihilation
cross sections for $\bar p$ on the same targets as used in the 
$\bar n$ experiment \cite{ABB02}.

\section{Antineutron-nucleus annihilation cross sections}
\label{sec:nbarsNUC}

The optical potential derived above for $\bar p$-nucleus interaction
was used to calculate annihilation cross section for $\bar p$
and $\bar n$ on the six targets at the seven energies studied by
Astrua et al. \cite{ABB02}. Applying the $\bar p$ potential
also to calculate annihilation cross sections of $\bar n$ is
justified in this energy range because no trace of 
isovector-dependence was observed neither in the antinucleon
annihilation on the proton (Sect.~\ref{sec:ptarget}) nor in the
optical potential for antiprotonic atoms \cite{FGM05}. Moreover,
total $\bar pp$ cross sections do not show any sign of 
structure up to 420~MeV/c \cite{BHC87}.

\begin{figure}
  \includegraphics[width=0.75\textwidth]{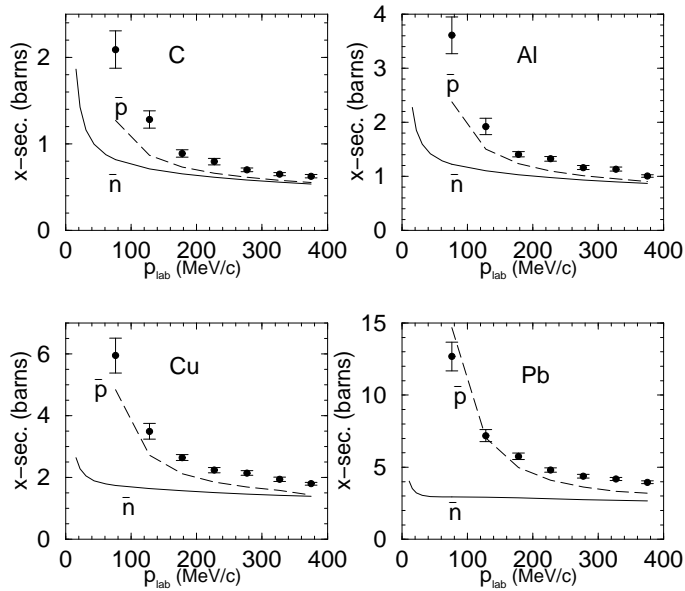}
\caption{Comparing calculation with experiment for total annihilation
cross sections  on nuclei using the potential
derived in Sect.~\ref{sec:pbarsNUC}. Solid curves for $\bar n$, dashed
curves for $\bar p$. Experimental results for  $\bar n$ are
from \cite{ABB02}. See also \cite{Fri14}.}
\label{fig:nbarsA}      
\end{figure}

Figure \ref{fig:nbarsA} compares calculated cross sections for $\bar p$
and $\bar n$ with experimental annihilation cross sections for $\bar n$
on nuclei.  The start of the $1/v$ rise for $\bar n$ at the lower momenta
is seen, shifting to lower and lower momenta as the size of the
nucleus increases. The calculated $\bar p$ cross sections increase
sharply at low momenta compared to the {\it calculated} cross sections
for $\bar n$, as expected. In fact, the momenta where the ratio
of the calculated $\bar p$ to $\bar n$ cross sections is 2
are in remarkable agreement with table \ref{tab:Coul}, showing
the dominance of Coulomb focusing at low momenta. Surprizingly, the
experimental cross sections for the $\bar n$ cross sections follow
closely the corresponding calculated cross sections for antiprotons.
In particular, the sharp rise of the experimental cross sections 
for antineutrons at
low momenta follows remarkably well the calculated cross sections
for antiprotons. This is in sharp contrast with the experimental
results for a proton target, as is seen in Fig.~\ref{fig:pbarnbarp}.

\begin{figure}
  \includegraphics[width=0.75\textwidth]{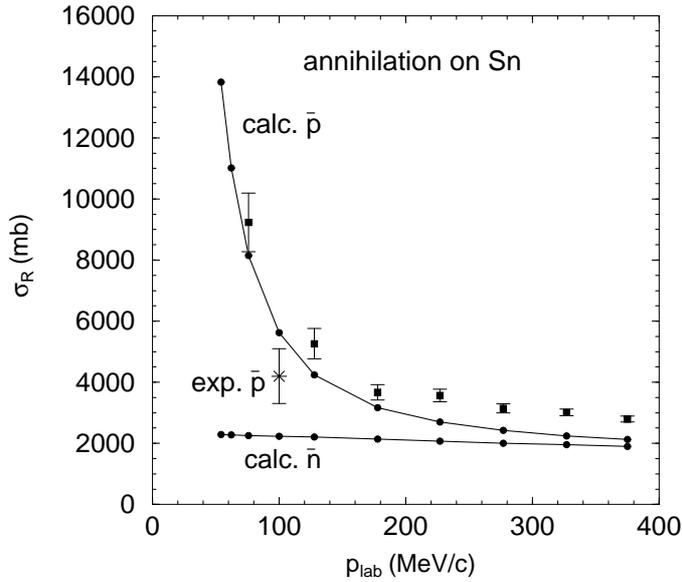}
\caption{Comparing calculation with experiment 
for total annihilation
cross sections on Sn using the potential
derived in Sect.~\ref{sec:pbarsNUC}. 
 Experimental results for  $\bar n$ (circles) are
from \cite{ABB02} and a single experimental result for $\bar p$
(star) is from \cite{BCH11}. This supersedes a similar figure 
in \cite{Fri14}.}
\label{fig:nbarsSn}      
\end{figure}

A direct comparison between experimental cross sections on nuclei
is possible only for a Sn target where a preliminary result is available
for $\bar p$ at 100~MeV/c \cite{BCH11}. Figure \ref{fig:nbarsSn} shows
the single measured $\bar p$ cross section and the seven 
experimental $\bar n$ cross sections in relation to the calculated
annihilation cross sections from the present optical potential.
The figure suggests that measured $\bar n$ cross sections around
100~MeV/c are larger than the corresponding $\bar p$ cross sections,
which is at variance with the results for a proton target as seen
in Fig.~\ref{fig:pbarnbarp}. Coulomb focusing (table \ref{tab:Coul})
suggests that the 
$\bar n$ annihilation cross sections on Sn are expected to be  
a factor of 3 {\it smaller} than the 
corresponding $\bar p$ cross sections near 100~MeV/c.

\section{Summary}
\label{sec:summ}
Total annihilation cross sections for antiprotons and antineutron
on the proton show smooth variation with energy between lab
momenta of 50 and 400~MeV/c.  The marked increase of the $\bar p$
cross sections relative to the $\bar n$ ones below 200~MeV/c
is fully consistent with the effect of Coulomb focusing. Above
that momentum the cross sections for the two types of projectile
are practically identical, suggesting that annihilation cross sections
at low energies are insensitive to isospin effects. Phenomenological
optical potentials reproduce well all the available results for
$\bar p$-nucleus interaction from antiprotonic atoms to elastic
scattering and annihilation cross section up to 600~MeV/c, with no
evidence for an isovector term. Consequently this potential is used 
to calculate annihilation cross sections, and comparisons with the
measured $\bar n$-nucleus cross section reveal
unexpected features of Coulomb effects. No explanation for this result
is known at present.

Direct comparison between
experimental cross sections without the help of a model is currently
possible only for a single measurement on Sn (in addition to the
many results for a proton target) and that shows
the $\bar n$ cross section to be larger than the $\bar p$ one. 
An experimental approach to this
`puzzle' may be possible in the foreseeable
future  by measuring total annihilation cross sections
for antiprotons on the six nuclear targets of Astrua et al. \cite{ABB02}
and at the same energies.

\begin{acknowledgements}
I wish to thank T. Bressani for useful discussions and for providing
in numerical form the detailed results of Ref. \cite{BBC97}.
Discussions with A.~Gal and A. Leviatan are gratefully acknowledged.
\end{acknowledgements}



\end{document}